\documentclass[12pt]{article}

\newtheorem{thm}{Theorem}

\newtheorem{prop}[thm]{Proposition}

\def\qed{{\bf QED}}

\def\be{\begin{eqnarray}}
\def\ee{\end{eqnarray}}
\def\bee{\begin{eqnarray*}}
\def\eee{\end{eqnarray*}}

\def\ts{\textstyle}

\def\rt2{\ts \frac{1}{\sqrt{2}} }

\title{Quantum Guessing via Deutsch-Jozsa}

\author{Michael Nathanson
\\ Department of Mathematics
\\ Northeastern University
\\ Boston MA 02115
\\ {\normalsize nathanson.m@neu.edu}
}

\begin{document}

\maketitle

\begin{abstract}
We examine the ``Guessing Secrets" problem arising in internet routing, in which the goal is to discover two or more objects from a known finite set. We propose a quantum algorithm using $O(1)$ calls to an $O(logN)$ oracle. This improves upon the best known classical result, which uses $O(logN)$ questions and requires an additional $O(logN^3)$ steps to produce the answer. In showing the possibilities of this algorithm, we extend the types of questions and function oracles that the Deutsch-Jozsa algorithm can be used to solve.
\end{abstract}

\pagebreak



\section{Introduction}
Inspired by challenges in internet routing \cite{P}, Chung, Graham, and Leighton proposed the ``Guessing Secrets" problem \cite{CGL}
in which a set of $k$ objects is to be reconstructed based on the
answers to a set of questions. It is a property of the answer
set as a whole that gives us information, suggesting that perhaps
a quantum algorithm could efficiently solve this problem. We propose using the Deutsch-Jozsa algorithm. Bernstein and Vazirani \cite{BV} showed a problem that requires $logN$ classical function calls but that is solved in one application of Deutsch-Jozsa; generalizations and other applications have followed \cite{BHH,CKL,CS,TS}. The present problem for two objects can be solved classically in $O(logN +
(logN)^3)$ steps \cite{AGKS}. In the quantum case, we can eliminate the $logN^3$ term of the complexity.

\medskip

The next section gives necessary background on the classical problem. It is followed by the presentation and analysis of the quantum algorithm in the case $k=2$. Finally, we present some explorations in a related problem with more objects.

\section{The problem for $k=2$ and its classical approaches}

\medskip

\noindent The following summarizes the basic problem and approaches for $k=2$ as presented by Chung, Graham, and Leighton \cite{CGL}:

\medskip

We are given a finite set of objects $\Omega$, of size $N$. An honest but possibly
malicious adversary selects elements $\lbrace
X_1, X_2 \rbrace \subset \Omega$. ÊOur task is to deduce as
much as possible about the $X_i$'s by asking yes/no
questions. A question is simply a map from $\Omega$ to $\lbrace 0,1
\rbrace$. For each question $q$, the adversary must respond with either $q (X_1)$ or $q(X_2)$. Our goal is to construct a set of questions $\lbrace q_j \rbrace$ and an
algorithm to use the responses to determine as much as
possible about $\lbrace X_1, X_2 \rbrace$.

\medskip

\noindent \textbf{Example:} Suppose $\Omega = \lbrace 1,2,3,4 \rbrace$. The Adversary chooses $\lbrace X_1, X_2 \rbrace =  \lbrace 1, 2 \rbrace$. If the question were ``Is your object an odd number?" the Adversary could answer on behalf of 1 and say ``Yes" or on behalf of 2 and say ``No." On the other hand, if the question were ``Is your object less than three?" the Adversary must answer ``Yes," since that answer is true for both objects in the set.

\medskip

There are inherent limits on what we can be assured of finding out. We can represent the elements of $\Omega$ as vertices of a graph; each question eliminates some number of possible pairs of objects, which correspond to edges in the graph. If our graph contains two disjoint edges $(X_1,X_2)$ and $(X_3,X_4)$, we can ask a question $q$ such that $q(X_1) = q(X_2) \ne q(X_3) = q(X_4)$, the response to which will eliminate one of the edges. There are only two types of graphs that contain no disjoint pairs of edges, a star and a triangle, and once we are in one of these configurations, the Adversary can prevent us from gaining any new information. 
A star, a set of edges all sharing a common vertex,
indicates that we know one of the objects we seek but
have limited information about the other; a triangle means that our two
objects are contained in a set of 3, but we don't know
which they are. The upshot is that the Adversary can limit our knowledge in
this way \textit{no matter what questions we ask}.
So, our goal is reduced to arriving at a point of maximal knowledge as
quickly as possible.
\medskip

Chung, Graham, and Leighton noted that the minimal size of the question set is $O(log N)$ whether the questions depend on
previous answers or not. As a result,
we will use only non-adaptive algorithms, where the
entire question set is submitted to the Adversary at
once. It is then reasonable to define a vector $A =
(A_1, A_2, \ldots, A_m)$, where $m$ is the number of questions asked and
$A_q$ is the adversary's response to question
$q$.
\medskip

Once we know the size of a minimal question set, the next goal is to create
questions that allow efficient implementation and
recovery of information about the $\lbrace X_i \rbrace$. One productive idea is to
represent $\Omega$ as $B^n$, a binary n-vector space
for $n = \lceil log_2 N \rceil$, and also to identify the questions $q$ as elements of
$B^n$, defining $q(X) = q \cdot X$, the inner product
(mod 2) of $q$ and $X$.

\medskip

Given a set of questions that generates maximal information,
efficiently extracting this information from the
responses is not easy. Both \cite{CGL} and \cite{AGKS} offer ways to do this, increasing
the size of the question set by a constant factor to
allow recovery in time polynomial in $logN$. In the quantum case, the
recovery of information is fundamentally intertwined with
the asking of the questions, so no distinction is made between them.

\medskip
In the classical algorithm, the number of questions is equal to the number
of calls to the adversary, and thus minimizing the
size of the question set is paramount. In the quantum model, we can
ask all the questions in superposition, so the number
of calls to the adversary is the more important measure of efficiency. We have chosen to
use all of $B^n$ for our question set, which, though much larger than the minimum necessary, has two
distinct advantages. First, it exhibits a symmetry
that we will take advantage of. Second, we can efficiently generate a
superposition of questions by applying
a Hadamard gate to each individual qubit; no multi-qubit operations are needed.
And while the set of questions is big, the number
of calls to the Adversary will be small.

\section{Quantum Procedure using Deutsch-Jozsa}

\subsection{Quantum Reformulations}
In their paper, Chung, Graham, and Leighton describe their problem as a generalization of traditional ``20 Questions", a game in which there is one unknown object $X \in \Omega$:

\medskip

\noindent \textbf{20 Questions}

Given an Adversary vector $A$ such that $\forall q, A_q = q(X)$ for some fixed $X$; find $X$.

\medskip

We compare this to the parity problem of Bernstein and Vazirani \cite{BV}, in which the quantum oracle inputs $q$ and returns the parity of the sum of a subset of the binary digits of $q$. This parity is exactly $q \cdot X$ (mod 2) for some $X$, so we can write:
\medskip

\noindent \textbf{Parity Problem:}

Given a quantum oracle for a function $f$, such that $\forall q, f(q) = q \cdot X$ for some fixed $X$; find $X$.

\medskip
Since we have defined our questions in terms of the dot product, it is clear that the parity problem is the quantum analogue of 20 Questions. In a similar way, we can formulate a quantum version of Chung, Graham, and Leighton's problem:

\medskip

\noindent \textbf{Quantum Guessing Secrets:}

Given a quantum oracle for a function $f$, such that $\forall q, f(q) = q \cdot X_1 \mbox{ or } f(q) = q \cdot X_2$ for some fixed $X_1, X_2$; find $X_1$ and $X_2$.

\medskip

In this formulation, we define a quantum oracle based
on the function $f(q) := A_q$, where $A$ is the vector of adversary responses described
above. This oracle will be a unitary map that takes $\vert q \rangle \vert
y \rangle$ to $\vert q \rangle \vert y \oplus A_q
\rangle$ with addition taken modulo 2. (Here, $\vert y \rangle$ represents
a single qubit, while $\vert q \rangle$ is an
n-qubit binary string $\vert q_1 q_2 \ldots q_n \rangle$ with $q = \sum_{i
= 1}^n q_i 2^{n-i}$). 

\medskip

The Deutsch-Jozsa algorithm was designed to distinguish between constant and balanced functions. A more careful inspection of
the measurement output shows that it can distinguish the $N$ different
functions $f_j \in ÊB^n$ defined by $f_j (X) = j \cdot
X$ \cite{J} and thus solves the parity problem \cite{BV}. We see then that the Deutsch-Jozsa algorithm solves ``20 Questions" in only one call to the oracle. This inspires us to apply it to Guessing Secrets.
\medskip

Much other work has been done to extend the classes of functions that this type of algorithm can distinguish \cite{BHH,CKL,CS}. In this paper we extend Deutsch-Jozsa in yet another direction by applying it to the guessing secrets problem. Terhal and Smolin \cite{TS} applied variations of Deutsch-Jozsa in order to recover an unknown object in a single query, but one oracle call will be insufficient for our purposes. There are $\left( \small{\begin{array}{c}N \\ 3 \end{array} }\right) + N ~$  triangles and maximal stars that our algorithm must distinguish, and Farhi, et al., have shown that such a quantum search must take at least 3 queries \cite{FGGS}.

\medskip

\subsection{The Quantum Algorithm}

\begin{prop}\label{prop1}
With a single call to the Adversary oracle, the Deutsch-Jozsa algorithm outputs one of the desired objects with probability at least one half.
\end{prop}

\noindent \textbf{Proof:} The algorithm is as follows:

\begin{itemize}
\item{Step 1:} Initialize $n$ bits to $\vert 0 \rangle$ plus an extra bit
to $\vert 1 \rangle$.
\item{Step 2:} Apply a Hamamard gate to each of the first $n$ bits, to
get a uniform superposition of all questions $q$. Also apply
a Hadamard gate to the final qubit to get an eigenstate of the
\textsc{NOT} gate.
\item{Step 3:} Apply the Adversary oracle.
\item{Step 4:} Apply a Hadamard gate to each of the first $n$ bits.
\item{Step 5:} Measure the first $n$ bits in the computational basis.
\end{itemize}

\noindent Formally, the system evolves like this:

\medskip

$\begin{array}{rl}
\vert 0 \rangle^{\otimes n} \vert 1 \rangle &  \stackrel{H^{\otimes n+1}}{\longrightarrow} 
\frac{1}{\sqrt{2N}}\displaystyle{\sum_{q = 0}^{N-1}}\vert q \rangle (\vert 0
\rangle - \vert 1 \rangle \nonumber \\
 & \stackrel{oracle}{\longrightarrow} \frac{1}{\sqrt{2N}}\displaystyle{\sum_{q = 0}^{N-1}} (-1)^{f(q)} \vert q \rangle (\vert 0
\rangle - \vert 1 \rangle) \nonumber \\
 & \stackrel{H^{\otimes n} \otimes I}{\longrightarrow}
\frac{1}{N } \displaystyle{\sum_{j = 0}^{N-1}} C_j\vert j \rangle ~ \frac{\vert 0 \rangle
- \vert 1 \rangle}{\sqrt{2}}
\end{array}$

\medskip

\noindent The coefficient of $\vert j \rangle$ is given by 

\be
C_j = \sum_{q=0}^{N-1} (-1)^{j
\cdot q + f(q)} = \# \lbrace q : ~ j \cdot
q = f(q) \rbrace - \# \lbrace q : ~ j \cdot
q \ne f(q) \rbrace.
\ee  

\medskip

What do we know about the coefficients of $\vert X_1 \rangle$ and $\vert X_2 \rangle$ (denoted $C_{X_1}$ and $C_{X_2}$)? For all $q$, $f(q) = q \cdot
X_1$ or $f(q) = q \cdot X_2$, so there
are only three types of questions $q$:

\begin{itemize}
\item{1:} $q \cdot X_1 \ne q \cdot X_2$ and $f(q) = q \cdot X_1$
\item{2:} $q \cdot X_1 \ne q \cdot X_2$ and $f(q) = q \cdot X_2$
\item{3:} $q \cdot X_1 = q \cdot X_2 = f(q)$
\end{itemize}
\noindent Let $S_i$ be the number of questions of type $i$, for $i = 1,2,3$.

\medskip

By the symmetry of $B^n$, for any distinct objects $X,Y
\in \Omega$, $q \cdot X = q \cdot Y$ for exactly
half the possible values of $q$. In particular $S_3 = N/2$. This allows us to calculate the probability that we measure $X_1$ or $X_2$:

\be
C_{X_1} & = & S_3 + S_1 - S_2 \nonumber \\
C_{X_2} & = & S_3 - S_1 + S_2 \nonumber \\
C_{X_1} + C_{X_2} & = & 2 S_3 ~  = ~ N \nonumber \\
C_{X_1}^2 + C_{X_2}^2 & \ge & \frac{N^2}{2}\nonumber \\
\left( \frac{C_{X_1}}{N}\right)^2 + \left( \frac{C_{X_2}}{N}\right)^2 & \ge & \frac{1}{2}
\ee

This means that the output of our measurement is one of the desired states with
probability at least $\frac{1}{2}$. \qed

\medskip

\subsection{Finishing the Procedure}
For given $\epsilon$, choose
values for $m$ and $d$ such that 
\bee
\mbox{Probability} \left( \mbox{Binomial}\left( m,\frac{1}{2}\right) \le \frac{m}{2} - d \right) < 
\epsilon
\eee
with $d$ as small as possible for the selected $m$. Run the algorithm $m$ times and let $E$ be the number of outputs equal to
either $X_1$ or $X_2.$  Then 
\bee
\mbox{Probability} \left( E \le \frac{m}{2} - d \right) < 
\epsilon
\eee
Let $F(X) =
$ the number of times $X$ appears as output in these
$m$ runs, for all $X \in \Omega$. Then with probability $1-\epsilon$, 
\be
(X_1, X_2) \in \left\lbrace (X^{\prime}, X^{\prime\prime}) : F(X^{\prime}) +
F(X^{\prime\prime}) \ge \frac{m}{2} - d \right\rbrace
\ee
There are two possible cases:
\begin{itemize}
\item{Case 1:} There is no $X \in \Omega$ with $F(X) \ge \frac{m}{2} - d$. Then
with probability $ 1 - \epsilon$, $\lbrace X_1, X_2
\rbrace$ is one of the edges $(X^{\prime}, X^{\prime\prime})$. The corresponding graph is not too complex, and complexity bounds depend only on $\epsilon$. We can then ask few questions sequentially of the oracle to reduce the graph to a triangle or star. 
\item{Case 2:} There is a dominant $X^{\prime}$ with $F(X^{\prime}) \ge
\frac{m}{2} - d$. Then all that can be said with probability
$ 1 - \epsilon$ is that the set of possible edges forms a star and that
$X^{\prime}$ is the center of it. This is
entirely analogous to the classical case, in which our remaining
uncertainty is determined by the number of points on the star.
While the classical algorithm was a method of elimination, the quantum
algorithm is a generator of outcomes; as such, it allows
us to discover the center of the star quite quickly, even if it makes it
harder to find all the possible second objects.
\end{itemize}

Let $X \in \Omega, X \notin \lbrace X_1, X_2\rbrace.$ For every run of the
algorithm, $C_{X_i} \ge \vert C_{X} \vert$ for $i =
1,2.$ (The requirement that half the answers for X match those for $X_1$
and $X_2$ ensures that $C_X$ can be neither too big
nor too small.) So the expected value $E\lbrack
F(X_i)\rbrack \ge E\lbrack F(X)\rbrack, i = 1,2.$ This fact will remain true even if the Adversary changes the oracle between calls and so will be
useful in any further statisical analysis of the outcomes.

\subsection{Examples}\label{exsect}
\begin{itemize}
\item{Full star:} The Adversary answers every question for the same object ($\forall ~ q ~ A_q = q \cdot X_1$). The
corresponding graph is a star centered at $X_1$ with $N-1$ points. The quantum
algorithm outputs $X_1$ with probability 1.
\item{Triangle:} The Adversary chooses a third object $X^*$ and answers for
it whenever there is a choice. (If $q \cdot X_1
\ne q \cdot X_2, A_q = q \cdot X^*$.) The corresponding graph is a triangle
on $X_1, X_2, X^*$. Our algorithm will output $X_1,
X_2, X^*, ~\mbox{or}~ X_1 \oplus X_2 \oplus X^*$, each with probability
$\frac{1}{4}$. At most three additional questions will be
needed to determine which 3 out of 4 are desirable.
\item{An intermediate case:} For questions when $q \cdot X_1 \ne q \cdot
X_2$, the Adversary chooses $q \cdot X_1$
three-fourths of the time. The corresponding graph is either a star
centered at $X_1$ with only a few points or simply the edge
$(X_1, X_2)$. The quantum algorithm outputs $X_1$ with probability
$\frac{9}{16}$, $X_2$ with probability $\frac{1}{16}$, and a
handful of others, each with probability $\leq \frac{1}{16}$.
\end{itemize}

\section{Comparing Complexities of the Classical and Quantum Algorithms}
The classical case for $k=2$ requires $O(logN)$ questions that are
answered sequentially. In the quantum case, each run of the algorithm requires only one call to the oracle. Since we are directly comparing the quantum performance to the classical, one must ask whether our large question set makes the oracle exponentially more complex. The answer is No. If the Adversary wishes to effect a triangle or an $(N-1)$-pointed star, it is straightforward to write an oracle algorithm in $O(logN)$ steps to assign appropriate values for $f(q)$. (More generally, a quantum adversary can force the guesser into a graph $G$ with an $O(logN)$ oracle if and only if a classical adversary could force the guesser into $G$ with any $O(logN)$ questions from the set $B^n$.) 
\medskip

Thus the questioning part of the algorithm takes $O(logN)$ steps in both cases. Classically, it is then necessary to extract information about $\lbrace X_1, X_2 \rbrace$ from the questions' answers; the list decoding method in \cite{AGKS} accomplishes this in $O((logN)^3)$ steps. ÊBy
contrast, the quantum algorithm simply runs $O(1)$ times and the output is in an immediately useable form. The number of runs
(and calls to the oracle) is \textit{independent}
of the size of $\Omega$ and depends only on the desired level of certainty. It is the avoidance of a complicated decoding stage that allows the quantum algorithm to perform faster than its classical analogue. 

\medskip

\section{What about $k > 2$?}

In the original paper, Chung, Graham, and Leighton begin with a more general problem in which the guesser tries to determine a set of $k$ objects, with the promise that for all questions $q$, the Adversary's response $A_q = q(X_i)$ for some $i \in \lbrace 1, 2, \ldots k \rbrace$. Both \cite{CGL} and \cite{AGKS} observe that the situation gets
much more complicated for $k > 2$, and the classical results in this case are far
from complete or satisfying. For $k = 2$, there are 2 shapes for the final graph (star and triangle); for $k = 3$, there are 8 final hypergraph shapes, and any algorithm must be able to distinguish $O(N^7)$ of them \cite{CGL}. According to \cite{FGGS}, a quantum algorithm that can differentiate these requires at least 7 oracle calls. However, our present algorithm
does not extend directly in the case of larger $k$. 

\medskip

To see this, let $k = 3$ and define the function by a ``minority rule" (for fixed $X_1, X_2, X_3, ~ f(q) := q \cdot X_1 ~ \oplus ~ q \cdot X_2 ~ \oplus ~ q \cdot X_3$); then $q \cdot X_i = f(q)$
for exactly half the questions and $C_{X_i} = 0$ for $i = 1,2,3$. The
algorithm will \textit{never} pick any of the correct
objects! This sort of difficulty will persist for all higher values of $k$. 

\medskip

It is possible to generalize the original problem differently by placing stronger restrictions on the function $f$. For example, suppose that when asked a question, the Adversary must respond on behalf of the majority of the $X_1, X_2, \ldots, X_k$. This example is formalized below: 

\medskip

\noindent \textbf{Majority Problem} 

Given a function $f: B^n \rightarrow \lbrace 0,1 \rbrace$ such that $\forall ~ q \in B^n, ~ f(q) = q \cdot X_i$ for \textit{at least half} of the $X_i \in \lbrace X_1, X_2, \ldots , X_k \rbrace$; determine $\lbrace X_1, X_2, \ldots , X_k \rbrace$. 

\medskip

Using the Deutsch-Jozsa algorithm with the function oracle \textit{does} solve this problem. As before, we define a run of the algorithm as successful if its output is one of the $X_i$. 

\medskip

\begin{prop}\label{propmax}

\begin{description}

\item{(i)} If the $X_i$'s are linearly independent as vectors in $B^n$, the algorithm succeeds with probability $\ge p_k$, where $p_k > \frac{2}{\pi}$ if $k$ is odd; $p_k \ge \frac{1}{2}$ if $k$ is even; and as $k \rightarrow \infty$, $p_k \rightarrow \frac{2}{\pi}$.

\item{(ii)} With no assumptions about the $X_i$'s and for $k$ odd, P(success) $\ge \frac{1}{k}$, and the order of this bound is strict. 

\item{(iii)} In the case $k=3$, the Deutsch-Jozsa algorithm solves the Majority Problem, producing one of the desired objects with probability $\ge \frac{3}{4}$.

\end{description}
\end{prop}

\noindent The issue of independence arises from the following: 

\medskip

\noindent \textbf{Symmetry Principle:} 

For all linearly independent sets of vectors $\lbrace X_1, X_2, \ldots, X_k \rbrace \subset B^n$ and $v \in B^k$, the vector $(q \cdot X_1, q \cdot X_2, \ldots, q \cdot X_k) = v$ for exactly $2^{n-k}$ questions $q \in B^n$. 

\medskip

It is obviously a disadvantage to require independence for our result. On the other hand, most sets of vectors $\lbrace X_1, X_2, \ldots, X_k \rbrace$ are linearly independent in the following sense: If $k$ vectors are chosen uniformly in $B^n$, a simple inductive argument shows that they are linearly independent with probability greater than $1 - 2^{k-n}$.
\medskip

\noindent \textbf{Combinatorial Proof of Proposition \ref{propmax}:}

\medskip

\noindent (i) For $M \subset \lbrace 1,2 \ldots, k \rbrace$, define 
\bee
\epsilon_{i,M} & := & \left\lbrace \begin{array}{l}
1 \mbox{ if } i \in M \\
-1 \mbox{ if } i \notin M
\end{array}\right. \\ 
S_M & := & \# \lbrace q \in B^n : ~ \lbrace i : ~ f(q) = q \cdot X_i \rbrace ~ = ~ M \rbrace
\eee

Thus $S_M$ counts the number of questions for which $M$ is the set of $X_i$ which agree with the function. 

\medskip

In Step 4 of the algorithm, the coefficient of $\vert X_i \rangle$ is $\frac{C_{X_i}}{N}$, where
\be
C_{X_i} & = & \sum_{
M \subset \lbrace 1,2, \ldots, k \rbrace}~ 
\epsilon_{i,M} ~  S_M. 
\ee

Observe that if $|M| < \frac{k}{2}$ then $S_M = 0$ by assumption, and if $|M| = \frac{k}{2}$ then $\sum_i ~ \epsilon_{i,M} = 0$. This means that once we sum over $i$, we can restrict our attention to subsets $M$ with $|M| > \frac{k}{2}$.
\be
\sum_{i=1}^k C_{X_i} & = & \sum_{i , M }~ 
\epsilon_{i,M} ~  S_M = \sum_{
i; |M| > \frac{k}{2}}~ 
\epsilon_{i,M} ~  S_M
\ee
If the $X_i$'s are linearly independent and $|M| > \frac{k}{2}$, then $S_M = 2 \left( 2^{n - k} \right) = \frac{N}{2^{k-1}}$ by the Symmetry Principle stated above, so  
\be
\sum_{i=1}^k C_{X_i}= \sum_{
i; |M| > \frac{k}{2}}~ 
\epsilon_{i,M} ~  S_M =  \frac{N}{2^{k-1}} \sum_{
i; |M| > \frac{k}{2}}~ 
\epsilon_{i,M}
\ee

\medskip

For a given $j$, there are $\left( \begin{array}{c} k \\  j \end{array}\right)$ subsets $M$ of size $j$. For a fixed $i$, we can then count how many $M$'s contain $i$ and how many do not. This is exactly the information encoded by $\epsilon_{i,M}$:
\be
\sum_{|M| = j} \epsilon_{i,M} = \left( \begin{array}{c} k - 1 \\  j - 1 \end{array} \right) - \left( \begin{array}{c} k - 1 \\  j \end{array} \right)
\ee
Summing over all $j > \frac{k}{2}$
\be
\sum_{i=1}^k C_{X_i}& = &\frac{N}{2^{k-1}} \sum_{i=1}^k \sum_{|M| > \frac{k}{2}}~ 
\epsilon_{i,M}\nonumber
\\ & = & \frac{N}{2^{k-1}} \sum_{i=1}^k \sum_{j= \lceil \frac{k+1}{2} \rceil}^k \left( \begin{array}{c} k - 1 \\  j - 1 \end{array} \right) - \left( \begin{array}{c} k - 1 \\  j \end{array} \right) \nonumber \\ & = &\frac{N}{2^{k-1}} \sum_{i=1}^k \left( \begin{array}{c} k - 1 \\  \lceil \frac{k+1}{2} \rceil -1\end{array} \right) \nonumber \\ & = &\frac{k N}{2^{k-1}}  \left( \begin{array}{c} k - 1 \\  \lceil \frac{k+1}{2} \rceil -1\end{array} \right)
\ee
Dividing by $N$ yields
\be
\sum_{i=1}^{k} \frac{C_{X_i}}{N} & = & \frac{k}{2^{k-1}} \left( \begin{array}{c} k - 1 \\  \lceil \frac{k-1}{2} \rceil \end{array} \right)
\ee
The sum of the squares of the coefficients is minimized if they are all the same.
\be
\mbox{P(success) }& = & \sum_{i=1}^k \left( \frac{C_{X_i}}{N}\right)^2 \ge \frac{k}{2^{2(k-1)}} \left( \begin{array}{c} k - 1 \\  \lceil \frac{k-1}{2} \rceil \end{array} \right)^2 =: p_k 
\ee

\medskip

It is easy to check that $p_2$ = $\frac{1}{2}$ and $p_3$ = $\frac{3}{4}$; and $p_{2m}$ increases monotonically to $\frac{2}{\pi}$ while $p_{2m + 1}$ decreases monotonically to $\frac{2}{\pi}$.
 
\medskip

\medskip

\noindent (ii) The assumption that $k$ is odd implies that for all $M \subset \lbrace 1,2, \ldots, k \rbrace$, either $|M| < \frac{k}{2}$ and $S_M = 0$, or else $(2|M| - k) \ge 1$. So for all $M$:
\be
(2|M| - k) S_M  \ge  S_M
\ee
Also, for any $M$, 
\be
\sum_i \epsilon_{i,M} = |M| - (k - |M|) =  2|M| - k
\ee
By definition, $\sum_M S_M$ is the total number of questions, so
\be
\sum_{i=1}^k C_{X_i} = \sum_{i,M} ~ 
\epsilon_{i,M} ~ S_M ~ = ~ \sum_M (2|M| - k) S_M  ~ \ge ~ \sum_M S_M  ~ = ~ N
\ee
Dividing by $N$ gives
\be
\sum_{i = 1}^k \frac{C_{X_i}}{N} & = & 1
\ee
The sum of the squares is minimized if each $\frac{C_{X_i}}{N} = \frac{1}{k}$.
\be
\mbox{P(success)} = \sum_{i=1}^k \left( \frac{C_{X_i}}{N}\right)^2 \ge k \left( \frac{1}{k} \right)^2 = \frac{1}{k}.
\ee

We cannot do significantly better than this bound, either. Let $G$ be a subgroup of $B^n$ and let $X_1, X_2, \ldots, X_k$ be the nonzero elements of $G$. (So $k$ is one less than a power of two; in particular, $k$ is odd.) Then $f(q) = 0$ if and only if $q \in G^\perp$, which implies that for all $i \in \lbrace 1,2,\ldots,k \rbrace$:
\be
C_{X_i}  = \left( \frac{N}{2} + \vert G^\perp \vert \right) - \left( \frac{N}{2} - \vert G^\perp \vert \right) = 2(\vert G^\perp \vert) = 2 \left( \frac{N}{k+1} \right)
\ee
Dividing by $N$ and summing the squares, we get
\be
\mbox{P(success)} & = & \sum_{i=1}^k \left( \frac{C_{X_i}}{N}\right)^2 = k \left( \frac{2}{k+1} \right)^2 = \frac{4k}{(k+1)^2}
\ee

This is $O(\frac{1}{k})$, so the bound above is more or less strict.

\bigskip
\bigskip

\noindent (iii) It should be noted that the Majority Problem for $k=3$ is equivalent to the earlier example of the triangle (Section \ref{exsect}), where it was stated without proof that $X_1, X_2,$ and $X^*$ would each appear with probability $\frac{1}{4}$. A complete proof follows immediately from the results (i) and (ii):

\medskip

Part (i) proved that $p_3 = \frac{3}{4}$ if the $\lbrace X_i \rbrace$ are independent. If they are dependent and all nonzero, then they are the nonzero vectors of a subgroup $G$ as described in (ii) and P(success) $ = \frac{4k}{(k+1)^2}= \frac{3}{4}$. Finally, if one of the vectors \textit{ is } zero, the vectors span a subgroup of dimension $2$ and it is straightforward to show that each vector in this group will appear with probability $\frac{1}{4}$. Thus in any case, the algorithm succeeds will probability $\ge \frac{3}{4}$. \qed

\medskip

\section{Conclusions}
The Deutsch-Jozsa algorithm was written to show the power of a quantum
computer in solving an artificially created problem. In
this paper, we applied the same algorithm in a different way to
an existing classical problem with active interest in combinatorics and computer science. This ``Guessing Secrets" problem can solved classically in $O(logN +
(logN)^3)$ steps while the quantum algorithm uses $O(1)$ calls to an $O(logN)$ oracle. The quantum algorithm achieves this higher efficiency by producing output in an immediately useable form. It is certainly worth considering what other sorts of problems might be addressed
in a similar fashion.

\bigskip
\noindent{\bf Acknowledgments:} ÊThis project was supported by
the GAANN program at Northeastern University.  I am grateful to Chris King for suggesting this problem and to both him and Beth Ruskai for their guidance and support.

{~~}

\end{document}